\documentstyle[prl,aps,epsf,12pt]{revtex}
\begin{document}
\def\inseps#1#2{\def\epsfsize##1##2{#2##1}\centerline{\epsfbox{#1}}}
\newcommand {\beqa}{\begin{eqnarray}}
\newcommand {\eeqa}{\end{eqnarray}}
\newcommand {\n}{\nonumber \\}
\newcommand {\beq}{\begin{equation}}
\newcommand {\eeq}{\end{equation}}
\newcommand {\om}{\omega}
\newcommand {\s}{\sigma}
\newcommand {\Si}{\Sigma}
\newcommand {\de}{\delta}
\newcommand {\pa}{\partial}
\newcommand {\e}{\epsilon}
\newcommand {\Lm}{\Lambda}
\newcommand {\th}{\theta}
\newcommand {\ga}{\gamma}
\newcommand {\bt}{\bigtriangleup}
\newcommand {\bd}{\Diamond}
\newcommand  {\cl}{\clubsuit}
\newcommand  {\ov}{\overbrace}
\newcommand  {\un}{\underbrace}

\title{
Power Laws are Logarithmic Boltzmann Laws}
\author{
Moshe Levy and Sorin Solomon \thanks{ Email: shiki@astro.huji.ac.il,
 sorin@vms.huji.ac.il;
 http://shum.huji.ac.il/$\sim$sorin }\\
Racah Institute of Physics,
Hebrew University, Jerusalem 91904, Israel\\
}
\date{8 July 1996}
\maketitle
$\ \ $ \\
To appear in Int. J. Mod. Phys. C \\
The paper can be also viewed as \\
 http://shum.huji.ac.il/$\sim$sorin/shikipap/polobo.ps
$\ \ $ \\
$\ \ $ \\
\begin{abstract}
{\bf
Multiplicative random processes in
(not necessaryly equilibrium or steady state) stochastic systems
with many degrees of freedom
lead to Boltzmann distributions when the dynamics is expressed
in terms of the logarithm of the normalized elementary variables.
In terms of the original variables this gives a
 power-law distribution.
This mechanism implies certain relations between
the constraints of the system, the power of the distribution
and the dispersion law of the fluctuations.
These predictions are validated by
Monte Carlo simulations and experimental data.
We speculate that stochastic multiplicative dynamics
might be the natural origin for the emergence of criticality
and scale hierarchies without fine-tuning.
}
\end {abstract}


In the last years researchers have found
an exceedingly large number of power laws
in very many natural and artificial (social, economic)
systems.

The emergence of "scaling" properties
was considered intriguing as in theoretically
known models this is related usually with very special
"critical" conditions.
In the parameter space of typical equilibrium statistical models,
critical systems correspond to subspaces of measure zero.
Yet scaling systems seem to show up in nature
much more often than this theoretically expected measure zero
abundance.
This lead researchers to coin the term self-organized criticality
(\cite{bak1}, \cite{bak2}).

In the present note we present a simple yet very general
explanation of the emergence of power laws.
According to our analysis, power-like systems are expected
to arise as naturally  as the Boltzmann distribution.

In fact we show that for a very large class of systems,
their power law distribution is in a precise mathematical
relation to a Boltzmann distribution when the measurables
are represented on a logarithmic scale.
This analysis implies additional relations which are confirmed
 experimentally.

Consider a system consisting of a large set of
elements $i$ which are characterized each by
a time-dependent variable $\omega_i (t)$
(for definiteness one can think of a set of investors
$i=1,...,N$ each owning a wealth $\omega_i$
or N towns containing each $\omega_i$ people).

Assume that the typical variations of $\omega$ are characterized
effectively by a multiplicative stochastic law:
\begin{equation} \omega_i (t+1) = \lambda \omega_i (t)
\label{mult}
 \end{equation}
with $\nu$ being a stochastic variable
with a finite support distribution of probability $\pi (\lambda )$.

The effective "transition probability" distribution $\pi (\lambda )$
is assumed not to depend on $i$ or on the actual value of $\omega_i$.
However, we will see that our conclusions are not affected
if the shape of $\pi (\lambda )$ varies in time during the process.

In order to isolate the {\bf shape} of the distribution of $\omega $
even for situations in which there is an unbounded
overall drift of the $\omega_i (t)$'s towards infinity, we will work
in the sequel of the article with the distribution ${ P}(w)$:
which fulfills the master equation:
\begin{equation}  P(w,t+1) - P(w,t)=
     \int_{\lambda} \Pi ({\lambda}) P(w/{\lambda},t) d{\lambda}
    - P(w,t) \int_{\lambda} \Pi ({\lambda}) d{\lambda}
\label{multmast} \end{equation}
 where $w$'s are normalized $\omega's$ such as to fulfill at each time:
\begin{equation} \sum_i w_i (t) \equiv \int w P(w,t) dw  = N
\label{norm}
\end{equation}
i.e. in the wealth case one represents actually
the {\it relative} wealth of each investor.
Correspondingly, the transition probability distribution $\Pi (\lambda )$
for the new variables is related to $\pi (\omega )$ by a shift
in the argument. 

Moreover, one limits from below the allowed values of $w > w_0$
(in the wealth case this consists in subsidizing
individuals as not to fall below a certain poverty line $w_0$).
This implies appropriate changes in the transition probability for
$w_i$'s in the immediate neighborhood of $w_0$.

In order to extract the implications of the
dynamics (\ref{multmast}) it is convenient to
represent it  on the logarithmic scale
in terms of $x = ln w$ and $\mu = ln \lambda$.
The corresponding probability distributions $P$ and $\Pi$ become
in the new variables:
\begin{equation}
{\cal P} (x) = e^x P(e^{x})
\label{transf}
\end{equation}
and respectively
${\rho} (\mu) = e^{\mu} \Pi (e^{\mu})  $ .
In terms of ${\cal P}, x, \rho , \mu$,
the master equation (\ref{multmast}) becomes:
\begin{equation} {\cal P}(x,t+1) - {\cal P}(x,t)=
     \int_{\mu} \rho ({\mu}) {\cal P}(x-{\mu},t) d{\mu}
    - {\cal P} (x,t) \int_{\mu} \rho ({\mu}) d{\mu}
\label{addmast} \end{equation}

Not that this equation has the standard form of the
master equation for an usual Monte Carlo process.

The iteration of the equation (\ref{addmast}) for long time sequences
projects upon the eigenmode with the largest eigenvalue of the
time evolution operator:
\begin{equation}
\Omega_{\rho} {\cal P} (x) \equiv
\int_{\mu} \rho ({\mu}) {\cal P}(x-{\mu}) d{\mu}
    + {\cal P} (x) \Bigl ( 1- \int_{\mu} \rho ({\mu}) d{\mu} \Bigr )
\label{oper} \end{equation}
This in turn leads to an asymptotic distribution of
${\cal P}$ which fulfills an equation of the form:
\begin{equation}
 {\int_{\mu} \rho ({\mu}) {\cal P}(x-{\mu}) d{\mu} }  =
\Lambda {\cal P} (x)
\label{eigen} \end{equation}
Ignoring for the moment the boundary and finite size effects,
one can
easily verify that the solution of this equation is:

\begin{equation} {\cal P}(x) \sim e^{-x/T}
\label{bolt}
 \end{equation}
with $T$ determined by the condition
\begin{equation}
 \int_{\mu} e^{{\mu }\over T} \rho (\mu) d\mu  = \Lambda
\label{choq}
\end{equation}
The uniqueness of the solution (\ref{bolt}), (\ref{choq})
is insured by the normalization condition (\ref{norm}),
the positivity  of the density distribution ${\cal P}$ and
by the fact that for positive $\rho$ the left hand side in
(\ref{choq}) is a convex function in ${1\over T}$.
A rigorous proof that the equation (\ref{eigen})
leads to (\ref{bolt}) is given in \cite{choquet}
and is based on the extremal
properties of the $G-harmonic$ functions on non-compact groups
(in our case the group of translations on {\bf R}).

When one translates back the exponential "Boltzmann" law
(\ref{bolt}) in terms of the original variables $w = e^x$
one gets according (\ref{transf}) a power-law distribution:
\begin{equation} P(w) \sim w^{(-1-1/T)}
\label{power}\end{equation}

If one ignores the departures from (\ref{bolt})
due to the (upper) boundary and finite size effects one can use the
normalization conditions for the total "wealth", $w$, eq. (\ref{norm})
\begin{equation} C \int_{w_0}^{\infty} w^{-{1\over T}} dw = N
\label{normw}\end{equation}
and for the total number of elements:
\begin{equation} C \int_{w_0}^{\infty} w^{-1-{1\over T}} dw = N
\label{normn}\end{equation}
in order to express $T$ in terms only of $w_0$:
\begin{equation}
T = 1- w_0
\label{temp}
\end{equation}
This power-law and the above relation
\footnote{For very low values of $w_0$ the finite size
effects and the upper bound cannot be ignored and equation
(\ref{temp}) is modified.
The modified relation is confirmed by Monte Carlo simulations too
\cite{ls1}.
In particular for $w_0 = 0$ one gets $T=\infty$.
}
are excellently confirmed by simulations \cite{ls1}
in various systems for a wide range of $w_0$'s and is
consistent with experimental data \cite{pareto}.

It appears therefore that $T$ is largely independent on the shape
of the transition probability distribution $\rho (\mu )$
(or $\Pi (\lambda ))$.
Physically, an intuitive understanding of this
result can be achieved by thinking of eq. (\ref{addmast})
in terms of a conservative system in which
an energy $\mu$ can be absorbed or emitted
by each degree of freedom $i$ according to the
("Monte Carlo") emission-absorption
probability distribution $\rho (\mu )$.

The emergence of a Boltzmann distribution is independent
on the details of the energy exchange mechanism:
it is more general than the details of the particular dynamical
process leading to it.
In fact, even if the process itself is not stationary and the
"transition probabilities" $\Pi (\lambda )$ and $\rho (\mu )$
depend on time, the distribution $P(w)$ can still converge:
modifying during the process (or during a Monte Carlo simulation)
the interactions from short range to infinite range from
2-body to  many-body from direct interactions to interactions
through the intermediary of a bath or of an "energy reservoir"
is known \cite{myclust}
not affect the Boltzmann distribution (\ref{bolt}).

One sees therefore that a power law is as natural and robust for a
stochastic multiplicative process as the Boltzmann
law is for an equilibrium statistical mechanics system.
Far from being an exception and requiring fine tuning
or sophisticated self-organizing mechanisms, this is
the default.

For our general mechanism to apply to a scaling system, the system
has to fulfill the effective  stochastic multiplicative law (\ref{mult}).
Yet, the mechanism by which each particular system is lead to fulfill
(\ref{mult}) might differ.
For instance in the towns example
this might be related with interactions
between town residents (residents moving upon marrying somebody from
another town, or upon entering a new employee-employer bound).
In large scale universe structures, 2-body gravitational
forces might lead to laws similar to (\ref{mult}).

In a series of papers we have studied in detail
theoretically and numerically
these mechanisms in the context of the stock market
(\cite{ls-1}, \cite{ls0}, \cite{ls1} ).

Our results turned out to explain in detail
the Pareto power law distribution of the individual incomes
(experimentally documented in \cite{atkinson})
as well as the L\'evy distribution in the market prices fluctuations
(reported in \cite{stanley}).

The Pareto law arises as a consequence
of eq. (\ref{power}) and of eq. (\ref{mult}).
These equations imply that the individual speculative incomes
$r_i (t) = \int (\lambda -1) \Pi (\lambda ) w_i (t) d\lambda$
are distributed by a power law (\ref{power}) too:
\begin{equation}
P(r) = {r}^{-1-{1\over {1-w_0} } }
\label{pareto}
\end{equation}
By a similar argument, one finds that the market price fluctuations
induced by individuals are also distributed according to the
(\ref{pareto}) law.
According the generalized central limit theorem,
a quantity which is a sum of random variables $r$ distributed
according to a probability distribution $r^{-\alpha}$
converges to the L\'evy distribution $L_{1-\alpha}$
of characteristic exponent $1-\alpha$.

Our analysis implies therefore (and the experimental available data
confirm) that if the individual wealth distribution is fitted by
a power law of exponent $-1-{1\over {1-w_0} } $ (Fig 1)
then the speculative income distribution
is governed by the same law (\ref{pareto}) (Fig 2)
and the market fluctuations are given by a
L\'evy distribution of characteristic exponent
$-{1\over {1-w_0} } $ Fig (3).

These relations are confirmed by the available experimental data
(and by the Monte Carlo simulation of microscopic
representations of the stock market \cite{ls1})
Figs. (1) (2) (3) with $- \alpha = -{1\over {1-w_0} }= -1.4$
\cite{new}.

We plan in a future publication to compare with experiment
the relations which our mechanism predicts between the
rate of inflation, the taxation policy and the lower income
bound (the poverty line $w_0$).

In the context of fundamental physics, one may hope that
the extension \cite {furst} of the result
(\ref{bolt})-(\ref{power}) to $G-harmonic$
distributions on general non-compact (Weyl gauge \cite{weyl})
groups $G$ might lead microscopic models naturally, without
fine tuning, to criticality and scale hierarchies.

Such discrete (lattice gauge) theories with non-compact group
might provide a unified context for treating
renormalization theory and time:
the continuous re-scaling (\ref{norm})
of the "running to infinity" degrees of freedom $w_i$
(\ref{mult}) suggests ${\lambda}^t$ as the microscopic
stochastic origin of both time flow and renormalization flow
(with the "extremal" \cite{furst} distribution $P$ as the fixed point).

{\bf Acknowledgments}

We thank D. Stauffer for intensive correspondence
on this and previous articles on the same subject.

We also acknowledge discussions with R.Iengo, G. Mack,
A. Schwimmer and partial support from the Germany-Israel
Foundation, from the Fund for Fundamental Research of the
Israeli Academy and from the S.A. Schonbrunn Research Endowment Fund.

\newpage

Figure Captions

\underbar{Figure 1:}

The distribution of wealth (log-log scale).
The solid line represents actual data from
Great Britain (Source: Inland Revenue Statistics, 1970)
Dots represent the wealth distribution in our simulations.
The dashed line is a
fit by a power law distribution with a slope of - 2.40
($\alpha = 1.4$).

\underbar{Figure 2:}

Empirical distribution of income (log-log scale).
Data from
Great Britain (Source: National Income and Expenditure 1970).
Dashed line represents
fit by a power law distribution with
slope of - 2.34.

\underbar{Figure 3:}

Distribution  of returns on the stock (semi-logarithmic scale).
The solid line represents
the L\'evy distribution with exponent $\alpha = 1.40$ ( scale
factor 0.00375).
Dots represent distribution in simulation. Diamonds represent
the empirical return distribution for the S\&P 500 index during 1984 - 1989
as reported by Mantegna \& Stanley \cite{stanley}. The dashed line
represents the Gaussian distribution with the empirical
standard deviation ($\sigma = 0.05$) .

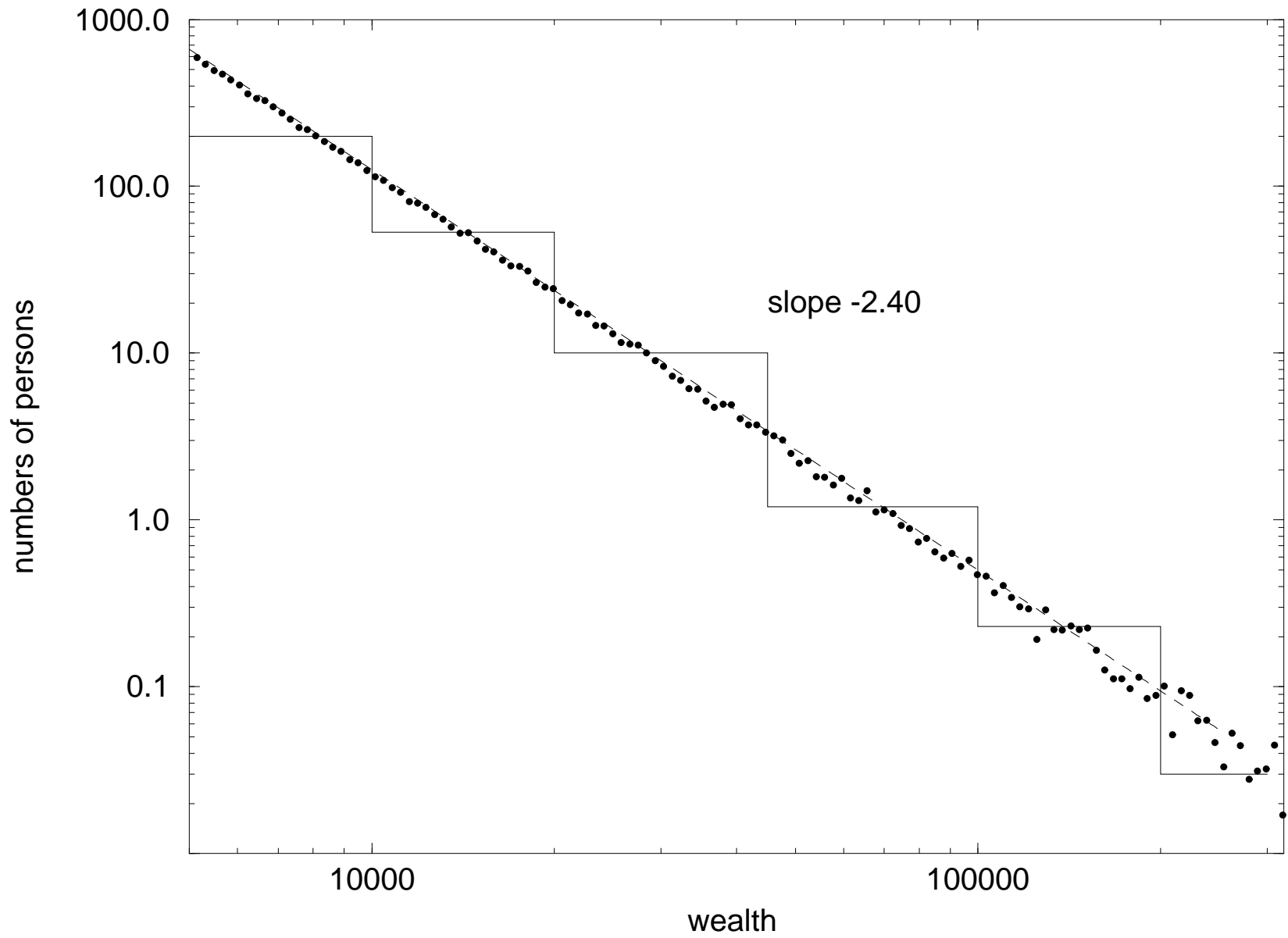
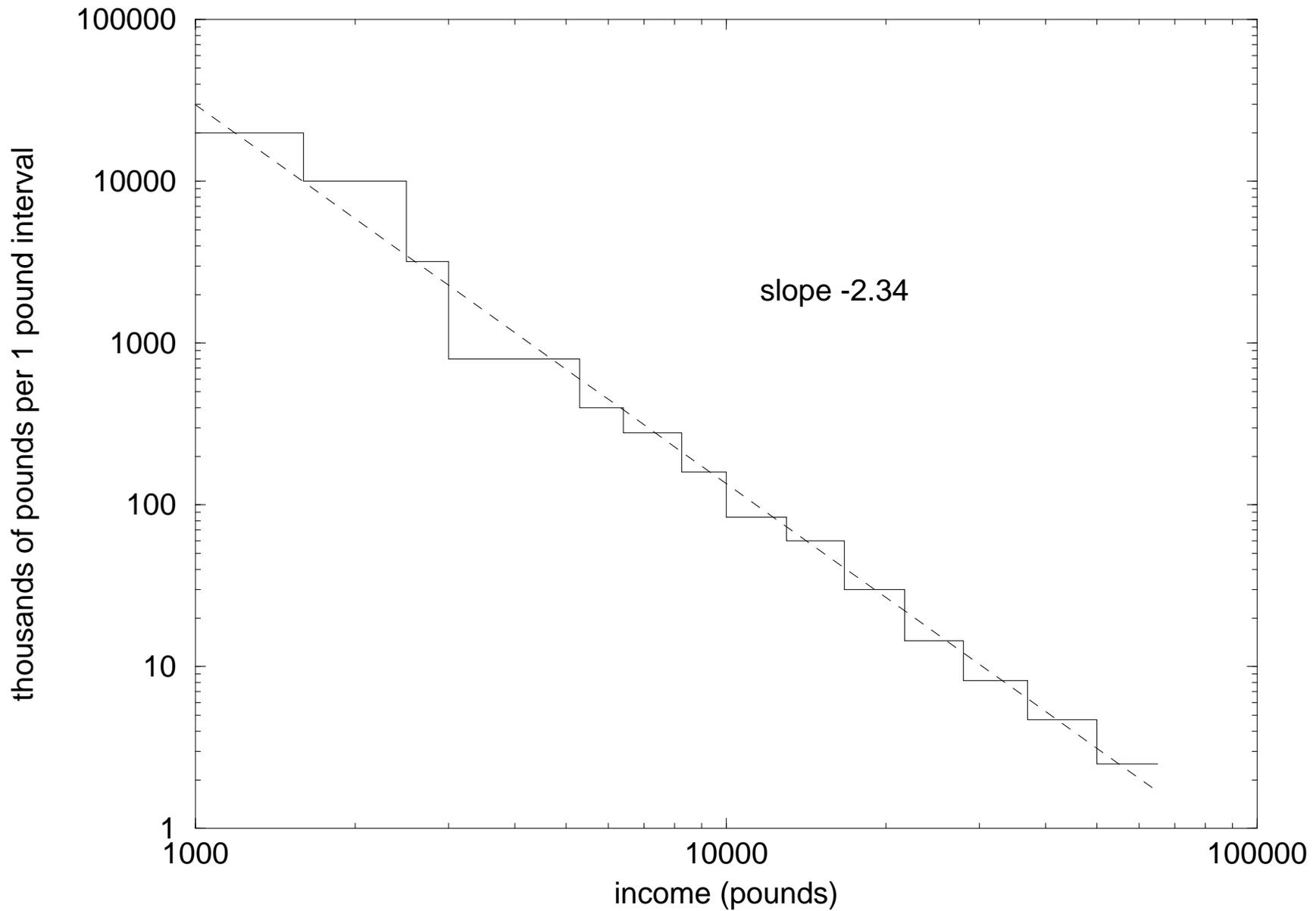
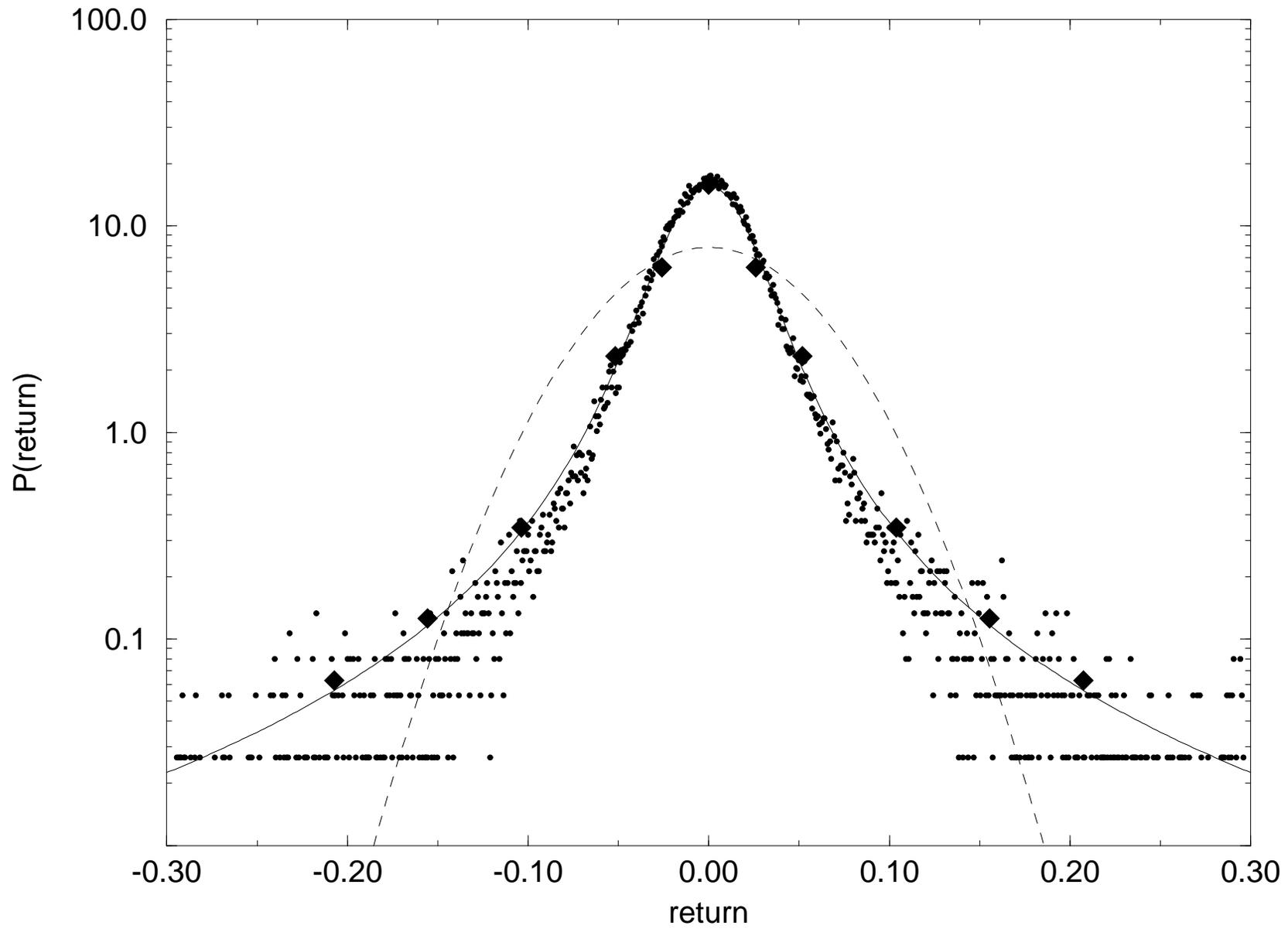
\end{document}